\documentclass[iop]{emulateapj}
\usepackage{apjfonts}
\usepackage{float}
\usepackage{times}
\usepackage{natbib}
\usepackage{rotating}
\makeatother 

\lefthead{Geach et al} 
\righthead{On the evolution of the molecular gas fraction}

\begin{document}

\title{On the evolution of the molecular gas fraction of star forming galaxies} 
\author{James~E.~Geach\altaffilmark{1}, Ian
Smail\altaffilmark{2}, Sean M.\ Moran\altaffilmark{3}, Lauren\ A.\
MacArthur\altaffilmark{4,5}, Claudia del P. Lagos\altaffilmark{2} \& Alastair
C. Edge\altaffilmark{2}}

\altaffiltext{1}{Department of Physics, McGill University, 3600 Rue
University, Montr\'eal, QC, H3A 2T8, Canada. jimgeach@physics.mcgill.ca}
\altaffiltext{2}{Institute for Computational Cosmology, Durham University,
South Road, Durham, DH1 3LE, UK} \altaffiltext{3}{Department of Physics and
Astronomy, Johns Hopkins University, Baltimore, MD 21218, USA }
\altaffiltext{4}{Herzberg Institute of Astrophysics, National Research Council
of Canada, 5071 West Saanich Road, Victoria, BC, V8X 4M6, Canada}
\altaffiltext{5}{Department of Physics \& Astronomy, University of Victoria,
Victoria, BC, V8P 1A1, Canada}

\submitted{Accepted version 16th Feb 2011}
\label{firstpage}

\begin{abstract}

We present IRAM Plateau de Bure interferometric detections of
CO\,($J$$=$$1$$\rightarrow$0) emission from a 24$\mu$m-selected sample of
star-forming galaxies at $z=0.4$. The galaxies have PAH\,7.7$\mu$m-derived
star formation rates of SFR$\sim$30--60\,$M_\odot$\,yr$^{-1}$ and stellar
masses $M_\star\sim10^{11}M_\odot$. The CO\,($J$$=$$1$$\rightarrow$0)
luminosities of the galaxies imply that the disks still contain a large
reservoir of molecular gas, contributing $\sim$20\% of the baryonic mass, but
have star-formation `efficiencies' similar to local quiescent disks and
gas-dominated disks at $z\sim1.5$--2. We reveal evidence that the average
molecular gas fraction has undergone strong evolution since $z\sim2$, with
$f_{\rm gas}\propto (1+z)^{\sim2\pm0.5}$. The evolution of $f_{\rm gas}$
encodes fundamental information about the relative depletion/replenishment of
molecular fuel in galaxies, and is expected to be a strong function of halo
mass. We show that the latest predictions for the evolution of the molecular
gas fraction in semi-analytic models of galaxy formation within a $\Lambda$CDM
Universe are supported by these new observations.

\end{abstract}

\section{Introduction}

Molecular hydrogen is arguably the most important component of the
interstellar medium (ISM) in high-redshift galaxies, since it is the phase
necessary for, and immediately preceding, star formation. However, observing
cool ($\lesssim$50\,K) molecular hydrogen is difficult; the molecule has no
permanent electric dipole and so one must rely on tracer molecular emission
(the most common being ro-vibrational emission from the $^{12}$CO isotopomer
[hereafter `CO']).

Unfortunately, uncertainty of the precise calibration of CO luminosity to
total gas mass in both low- and high-$z$ galaxies have made interpretation of
results challenging; even in the local Universe there is seen to be
metallicity and luminosity dependence on the conversion (Boissier et al.\
2003; Blitz et al.\ 2007; Komugi et al.\ 2010). Furthermore, although the bulk
of the molecular gas in star-forming disks is contained in virialised clouds,
it is clear that the most luminous galaxies -- mostly mergers -- have
molecular gas distributions and star-formation modes dramatically different to
quiescent disks; they require an alternative molecular mass calibration
related to the physical conditions of their interstellar media (Solomon et
al.\ 1997). The importance of understanding how the thermodynamic state of the
gas affects the calibration also remains controversial (high-$z$ observations
typically target $J_{\rm upper}>1$, requiring some correction to
$J=(1\rightarrow0)$ based on the excitation of the gas; although see Ivison et
al.\ 2010; Harris et al.\ 2010).

The majority of molecular gas studies at high-$z$ have, necessarily, focused
on the most active systems (generally sub-millimeter selected galaxies and
quasars), and so comparatively little is known about the evolution of the
molecular gas properties of the more common, but less active population.
Nevertheless, recent studies have started to make progress in expanding the
parameter space of CO observations at high-$z$, and several studies have
detected CO emission from `normal' disks at $z=1$--$2$ (Daddi et al.\ 2008;
2010; Tacconi et al.\ 2010). The striking result it that the efficiency of
star-formation in these distant galaxies is claimed to be similar to that seen
in local `quiescent' disks (Dannerbauer et al.\ 2009; Daddi et al.\ 2010),
with enhanced SFRs driven simply by larger gas reservoirs, drained by
star-formation (and feedback) on Gyr timescales. This hints that a secular
mode of star formation might be ubiquitous in $L^\star$ disks independent of
cosmic epoch (but see the higher-$z$ results of Coppin et al.\ 2007 and
Riechers et al.\ 2010).

The aim of this Letter is to further expand the parameter space of molecular
gas studies in normal galaxies, building on our observations of luminous
infrared galaxies (LIRGs) at $z=0.4$ (Geach et al.\ 2009, G09). We present new
IRAM Plateau de Bure interferometric (PdBI) observations of a sample of
24$\mu$m-selected galaxies in the outskirts of the rich cluster Cl\,0024+16 at
$z=0.395$. These galaxies bridge the gap between quiescent, local spirals and
luminous, high-redshift starbursts and disks. Throughout we adopt a
$\Omega_{\rm m}=0.3$, $\Omega_\Lambda=0.7$, $H_0=70$\,km\,s$^{-1}$\,Mpc$^{-1}$
cosmology.

\begin{table*} 
	\caption{Properties of the galaxy sample}\vspace{-0.5cm}
\begin{center}
\hspace{-0.1cm}
\begin{tabular*}{0.99\textwidth}{@{\extracolsep{\fill}}lccccccccccc}
  \hline
  Target & $\alpha_{\rm J2000}$ & $\delta_{\rm J2000}$ & $z_{\rm [O~{\sc II}]}$  &
  $L_{\rm FIR}$ &  SFR & $V^{\rm CO(1-0)}_{\rm FWHM}$&   $L'_{\rm CO(1-0)}$ & $M({\rm H_2+He})^{\dagger}$ & $M_\star^{\dagger\dagger}$ & $f_{\rm gas}$\cr 
  & (h\ m\ s) & ($\circ$\ $'$\ $''$) &  &
  ($10^{11}L_\odot$)  & ($M_\odot$\,yr$^{-1}$) & (km\,s$^{-1}$)  & (10$^{10}$\ K\ km\ s$^{-1}$\ pc$^2$)  & ($10^{10}M_\odot$) & ($10^{11}M_\odot$)\cr
  \hline
  
  MIPS J002652.5 & 00\ 26\ 52.5 & $+$17\ 13\ 59.9 & 0.3799 & $3.5\pm0.5$ &
  $62\pm19$ & $140\pm10$ &  $0.64\pm0.05$ & $2.9\pm0.2$ &0.95 & $0.23\pm0.02$ \cr
  MIPS\ J002703.6 & 00\ 27\ 03.6 & $+$17\ 11\ 27.9 & 0.3956 &  $2.3\pm0.3$&
  $42\pm12$ & $250\pm30$ &  $0.43\pm0.06$ & $2.0\pm0.3$ &0.87  & $0.19\pm0.03$\cr
  MIPS\ J002715.0 & 00\ 27\ 15.0 & $+$17\ 12\ 45.6 & 0.3813 & $1.9\pm0.3$ &
  $35\pm11$ & $340\pm40$ & $0.26\pm0.03$ & $1.2\pm0.3$ &1.10 & $0.10\pm0.03$\cr
MIPS J002609.1 & 00\ 26\ 09.1 & +17\ 15\ 11.5 &0.3940 & $1.5\pm0.3$ & $28\pm9$ & $\cdots$& $<0.39$ & $<1.8$ & 0.41 & $<0.30$\cr
MIPS J002606.1 & 00\ 26\ 06.1 & +17\ 04\ 16.4 & 0.3904& $1.9\pm0.3$ & $34\pm10$ & $\cdots$ & $<0.25$ & $<1.1$ & 0.49 & $<0.19$\cr

\hline
 MIPS\,J002621.7$^\star$ & 00\ 26\ 21.7 & +17\ 19\ 26.4 & 0.3803  & $3.1\pm0.2$ &
  $56\pm16$ & $140\pm10$  & $0.68\pm0.06$ & $3.1\pm0.3$ & 1.12 & $0.22\pm0.02$\cr
  MIPS\,J002721.0$^\star$ &  00\ 27\ 21.1 &  +16\ 59\ 49.9 & 0.3964  & $3.2\pm0.2$ &
  $59\pm16$  & $160\pm30$ & $1.14\pm0.11$ & $5.2\pm0.5$ & 0.98 & $0.34\pm0.04$\cr
  \hline \multicolumn{10}{l}{$\dagger$\,Assuming a CO luminosity to total gas
  mass conversion factor of $\alpha = 4.6\,M_\odot({\rm
  K\,km^{-1}\,pc^2})^{-1}$ that includes helium}\cr
  \multicolumn{10}{l}{$\dagger\dagger$\,Uncertainties $\sim$0.02\,dex based on
  refitting spectral templates 1000 times with random perturbation within
  1$\sigma$ photometric uncertainties.}\cr \multicolumn{10}{l}{$\star$\,Geach
  et al.\ (2009)} \cr
\end{tabular*}
\end{center} 
\end{table*}

\section{Observations and data reduction}

Our sample consists of five {\it Spitzer} MIPS 24$\mu$m detected LIRG-class
galaxies in the outskirts of the rich cluster Cl\,0024+16 ($z=0.395$, see
Geach et al.\ 2006). We obtained mid-infrared spectroscopy with {\it Spitzer}
IRS to confirm the presence of aromatic features, a clear indication of star
formation (rather than infrared emission dominated by an active galactic
nucleus; Geach et al.\ 2009b). Combined with the two galaxies observed with
IRAM PdBI in G09, this represents an effort to obtain CO detections or limits
for {\it all} the galaxies where we have detected significant emission from
the polycyclic aromatic hydrocarbon 7.7$\mu$m band. The line luminosities
imply star formation rates of $\sim$30--60\,$M_\odot$\,yr$^{-1}$, thus
allowing us to link molecular gas with moderate levels of dusty star formation
in the galaxies.

In this Letter we present the results in the context of the broader
star-forming population at $z=0.4$, rather than focusing on potential
environmental effects in the cluster. We believe this is a valid approach,
since these galaxies are expected to be randomly accreted onto the cluster
from the surrounding field, and the galaxies are seen at a stage where strong
environmental effects associated with clusters (specifically ram-pressure
stripping and harassment) are yet to have an effect (Treu et al.\ 2003; Moran
et al.\ 2007). They lie at clustocentric radii of 1.8--4.6\,Mpc
($\sim$1--3$\times$ the virial radius of Cl\,0024+16). Intrinsically, they are
likely to be representative of disk galaxies of equivalent mass sampled from a
random field, and can be compared to galaxies with similar properties at low
and high redshift; i.e.\ normal star forming disks.

The new IRAM observations were conducted over June--October 2009 in
configuration `D', using 5 antennae, and we adopted the same strategy as in
G09. Sensitivities ranged between 0.51--1.17\,mJy (median average for 10\,MHz
wide channels and 2 polarisations), and the on-source exposure times ranged
between 6.4--12.8\,hrs. We targeted the CO\,($J$$=$$1$$\rightarrow$0)
115.27\,GHz rotational transition at $\nu_{\rm obs} \simeq 82.63$\,GHz in 5
sources. The 3\,mm receiver was tuned to the frequency of the redshifted
CO\,($J$$=$$1$$\rightarrow$0) line at the systemic redshift of each galaxy
derived from optical spectroscopy (Czoske et al.\ 2000; Moran et al.\ 2007).
As in G09, the correlator was set-up with 2.5\,MHz spacing (2$\times$64
channels, 320\,MHz bandwidth). The phase and flux calibrators were the sources
3C454.3, 0119+115 and 0007+171. The observing conditions were good or
excellent in terms of atmospheric phase stability, however any anomalous and
high phase-noise visibilities were flagged in the calibration stage. Data were
calibrated, mapped and analysed using {\sc gildas} (Guilloteau \& Lucas\
2000).

\section{Results}

\subsection{Plateau de Bure CO detections}

We detect CO\,($J$$=$$1$$\rightarrow$0) emission in three galaxies
($>$4$\sigma$ detections within 2$''$ of the phase tracking centre,
significances determined from the integrated line flux). Velocity-integrated
maps and mm-spectra are shown in Figure\ 1. Total fluxes are evaluated from
single or double Gaussian fits to the spectra depending on whether the
profiles are very broad and have hints that they are double peaked. The line
luminosities are in the range $L'_{\rm
CO}=0.26$--$0.64\times10^{10}$\,K\,km\,s$^{-1}$\,pc$^2$ (Table\ 1).
Uncertainties on the luminosities are estimated by re-evalutating the Gaussian
fits after artificially adding noise to each channel based on the observed
r.m.s.\ fluctuations in each data cube. Upper 3$\sigma$ limits on $L'_{\rm
CO}$ for the two non-detections are based on the r.m.s. noise in the
observations of each source (note that the two non-detections are also the two
galaxies with the lowest SFRs in the sample). The luminosities and line widths
are listed in Table\ 1; we also list the properties of the two galaxies from
G09 which are included in the following analysis.

The spectra of MIPS\,J002652.5 and MIPS\,J002715.0, like MIPS\,J002721.0 (G09)
exhibit broad CO\,($J$$=$$1$$\rightarrow$0) emission, and can be well-fit by
double gaussian profiles. None of the new sample show obvious signs of major
mergers or strong tidal interaction in the deep optical imaging (Fig.\ 1), and
so we conclude that we are most likely observing CO\,($J$$=$$1$$\rightarrow$0)
emission tracing molecular gas distributed over rotationally-supported disks.

\begin{figure*}
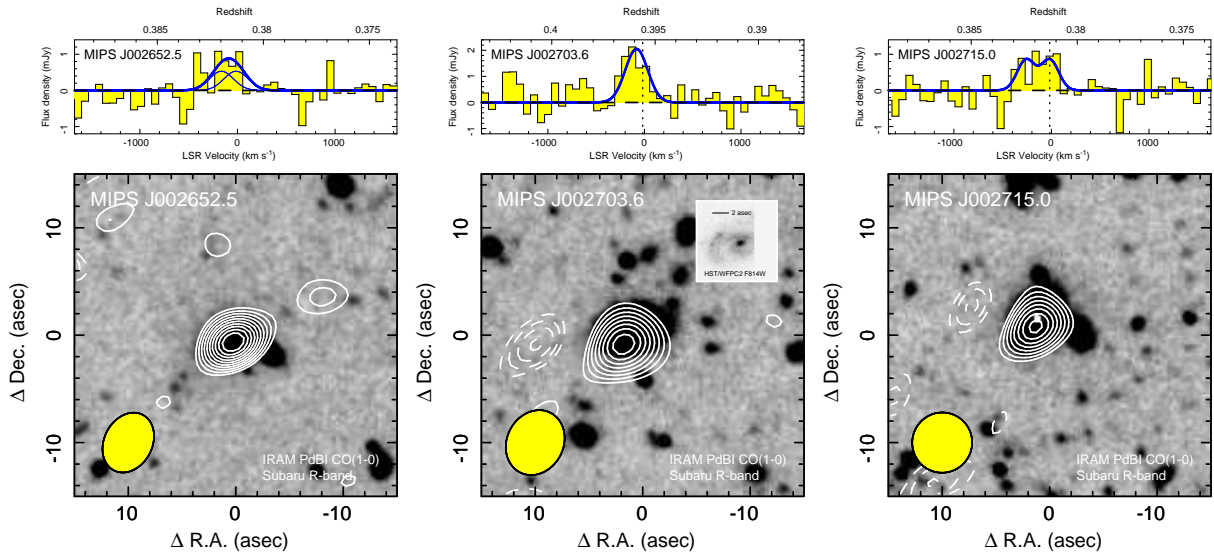

\centerline{\includegraphics[width=0.3\textwidth]{f1a.ps}\includegraphics[width=0.3\textwidth]{f1b.ps}\includegraphics[width=0.3\textwidth]{f1c.ps}}
\caption{Velocity integrated, clean CO\,$(J=1\rightarrow0)$ line emission maps
(at levels of $\geq$$2\times{\rm r.m.s.}$ in steps of 0.5) overlaid on
$30''\times 30''$ Subaru SuprimeCam $R$-band images (including a
$5''\times5''$ {\it HST}/WFPC2 inset for MIPS\ J002703.6). The ellipse
indicates the beam size and shape. ({\it top}) 3\,mm spectra extracted from
the peak of the integrated line emission maps, fit with either a single or
double Gaussian (where doubles have a fixed amplitude and width for each
peak).} \end{figure*}

\subsection{Physical properties of the galaxies}

Stellar masses for all the galaxies were determined by fitting the galaxy SED
constructed from {\it BVRIJK} imaging (Moran et al.\ 2007) to a large suite of
model SEDs using the {\sc Kcorrect} software, v4.2 (Blanton et al.\ 2007).
Utilizing the known spectroscopic redshifts, {\sc Kcorrect} finds the best
non-negative combination of the template spectra to fit the galaxy SED in its
rest-frame, including a reddening law, with best-fit values covering a range
$A_V=0.8-2.3$\,mag. Stellar masses are then calculated from the luminosity and
mass-to-light ratio of the best-fitting model in the rest-frame {\it K}-band,
$\Upsilon_\star=M_\odot/L^K_\odot$. The fits had an average mass-to-light
ratio of $\left< \Upsilon_\star\right> =0.36$ (range 0.24--0.52), and inferred
stellar masses ranging over $M_\star\sim$0.5--$1\times10^{11}M_\odot$ (Table\
1).

In G09 we adopted a conservative `ULIRG' conversion of
$\alpha=0.8$\,$M_\odot\,({\rm K\ km\ s^{-1}\ pc^2})^{-1}$ for gas mass, where
$M({\rm H_2+He}) = \alpha L_\odot$ (we omit the units of $\alpha$ for clarity
in the following). However, there are hints that a Galactic scaling is more
appropriate for these galaxies. The first point to note is the similarity
between the so-called star-formation `efficiencies', measured by $L'_{\rm
CO}/L_{\rm IR}$, compared to local quiescent disks. The disks in this sample
have a mean $\left< L_{\rm IR}/L'_{\rm CO}\right> = 51$ (range 28--73),
similar to that of local spirals, but lower than that of ULIRGs, which tend to
have $L_{\rm IR}/L'_{\rm CO} \gtrsim 100$ (Solomon et al.\ 1997). Daddi et
al.\ (2010) and Tacconi et al.\ (2010) report similar `secular' properties for
$z>1$ disks, although note that these high-$z$ observations measure $J_{\rm
up}>1$ transitions, and so require an additional uncertain correction to
estimate gas mass. It is claimed that the efficiency of star-formation remains
relatively constant in typical disks at all epochs (although, as a concept,
star-formation `efficiency' only makes sense if all of the molecular gas has
an equal probability of undergoing star-formation).

Although the CO (and mid-infrared) emission is unresolved, we can evaluate how
appropriate $\alpha=0.8$ is based on the offset from the Kennicutt-Schmidt
(K-S) law ($\Sigma_{\rm SFR}\propto\Sigma_{\rm gas}^{1.4}$; Kennicutt\ 1998),
assuming both gas and star-formation trace the stellar emission. Inclination
and effective radii of the disks are estimated by fitting a 2D model following
a S\'ersic profile to a CFHT {\it R}-band image (Czoske et al.\ 2001), using
{\sc galfit} v3.0.2 (Peng et al.\ 2010). {\sc galfit} was run on postage
stamps of $20''\times20''$, simultaneously fitting models to the main galaxy
and any bright companions in the field. {\sc galfit} produces estimates of the
galaxy axial ratio and size that accounts for the seeing of the ground based
image, by convolving with a point-spread function that we have extracted from
stars in the CFHT image. The effective radii of the disks range between
0.9--1.5$''$ (4.8--8.1\,kpc), with a mean of 1$''$ (5.4\,kpc), and typical
uncertainties of 0.02$''$. We take the effective radii as the size of the
disks and find a mean $\left< \alpha_{\rm KS}\right>\sim4.2$, in line with a
Galactic conversion. Note that this should be taken only as a guide for
choosing between a Galactic and ULIRG-like conversion, $\alpha_{\rm KS}$
should not be applied as the actual calibration: it is particularly sensitive
to the assumed spatial distribution of the gas and star-formation, such that a
larger assumed radius would result in a larger $\alpha_{\rm KS}$ and vice
versa (with $r\lesssim1$\,kpc required for $\alpha_{\rm KS}\sim0.8$).

MIPS\,J002703.6, is partially covered by one of the sparse {\it Hubble Space
Telescope}/WFPC2 (F814W) images taken of Cl\,0024+16 (Treu et al.\ 2003),
revealing finer detail of the optical morphology (Figure\ 1). The galaxy has
been typed as an early-type spiral (Treu et al.\ 2003), and has clear spiral
structure with several bright knots of emission in the northern arm
(presumably star-forming regions) and a potential bar. We have run {\sc
galfit} on the {\it HST}/WFPC2 image, where the best 2D fit is a two-component
S\'ersic bulge and exponential disk, where the bulge and disk have effective
radii of 0.15$''$ (0.8\,kpc) and 1.05$''$ (5.6\,kpc) respectively (consistent
with the value derived from the ground-based imaging), and the bulge-to-total
light ratio is $B/T=0.031$.

Based on this evidence, we now assume a Galactic conversion of $\alpha=4.6$ to
estimate the gas masses, which implies molecular gas masses of $M({\rm
H_2+He})=$1.2--$2.9\times10^{10}M_\odot$. What is the gas contribution to the
total baryonic mass of these galaxies? With our choice of $\alpha$, the
resulting gas fractions, $f_{\rm gas}=M_{\rm gas}/(M_{\rm gas} +M_\star)$, are
in the range 0.10--0.34, with a mean of $\left<f_{\rm
gas}\right>=0.21\pm0.03$. We can conclude that these are actively star-forming
galaxies with large gas reservoirs, but the disks are dominated by stellar
mass. This is markedly different to the recently discovered $z\sim1$--$2$
gas-rich disks of Daddi et al.\ (2010) and Tacconi et al.\ (2010), which are
generally more active galaxies, with comparable stellar masses and higher
gas-fractions $f_{\rm gas}\gtrsim40$\%.

\begin{figure*}
\centerline{\includegraphics[width=0.6\textwidth,angle=-90]{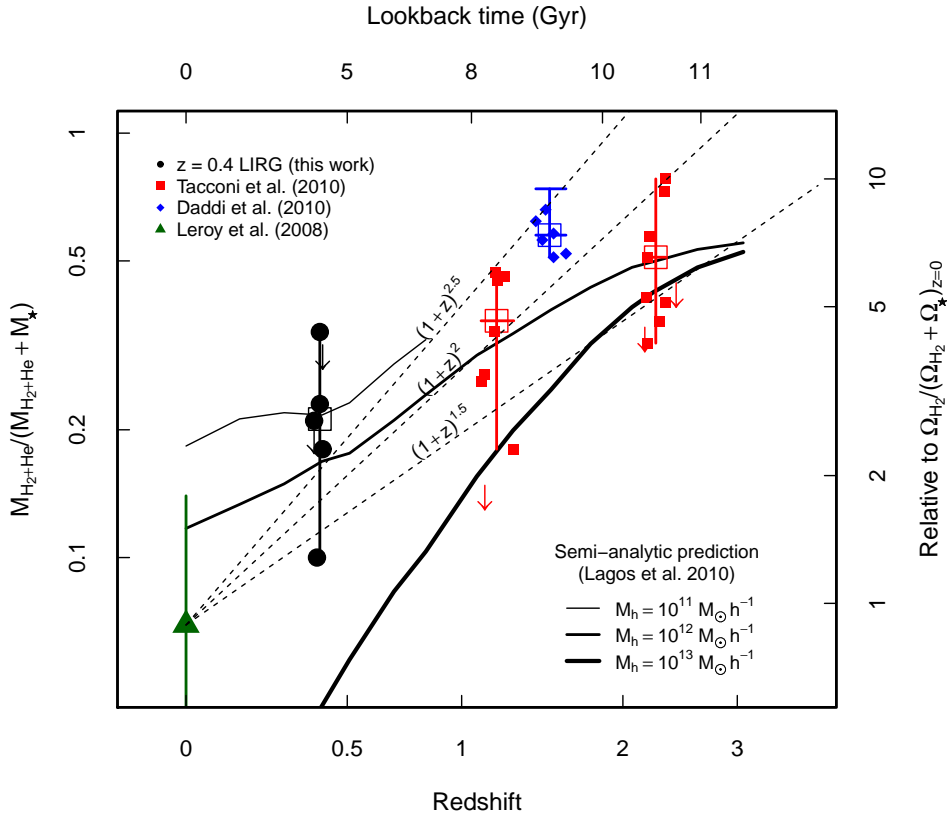}}
\caption{Evolution of the molecular gas fraction in $M_\star>10^{10}M_\odot$
galaxies, showing decline in $f_{\rm gas}$ over cosmic time following
$(1+z)^{\sim2\pm0.5}$. We compare our $z=0.4$ results with the high-$z$ disks
of Daddi et al.\ (2010) and Tacconi et al.\ (2010), and the $z\sim0$ sample of
Leroy et al.\ (2008). Large open symbols show the average of each sample, with
error-bars indicating the range. The upper bar of the Daddi et al.\ sample
shows the effect of assuming $\alpha=4.6$. All stellar mass estimates assume a
Chabrier initial mass function (IMF; we have applied a factor 1.25 to the
Leroy et al. data). We also show the evolution of $f_{\rm gas}$ in galaxies
with $M_\star \geq 10^{10}M_\odot$ in different halos from the latest
semi-analytic models (Lagos et al.\ 2010).} \end{figure*}

\section{Interpretation}

Figure\ 2 shows the observed $f_{\rm gas}$ in $M_\star\geq10^{10}M_\odot$
galaxies at $z=0$ (Leroy et al.\ 2008), $z=0.4$ (this work) and $z=1.5$--2
(Daddi et al.\ 2010, Tacconi et al.\ 2010) showing the clear decline in
$\left< f_{\rm gas}\right>$ in the 10\,Gyr since $z\sim2$. As a guide, we cast
the evolution of $f_{\rm gas}$ relative to the cosmological abundance of
baryons in stars and molecular gas at $z=0$: $\Omega_{\rm H_2}/(\Omega_{\rm
H_2}+\Omega_\star) = 0.078^{+0.160}_{-0.039}$ from the `baryon census' of
Fukugita, Hogan \& Peebles (1998). At $z=0$, the observed $f_{\rm gas}$ in
star-forming disks is close to that inferred from the total relative abundance
of molecular gas and stars in the Universe, but at $z\sim2$ it was a factor
$\sim$5--10 larger than the local value\footnote{The similarity of the average
galaxy gas fraction and the cosmic abundance at $z=0$ is perhaps coincidental,
since $\Omega_\star$ is integrated over the luminosity functions of all
bulges, disks and irregulars, whereas $\Omega_{\rm H_2}$ is inferred from a
relatively limited local CO survey. However, it could be a convenient datum
for comparisons of $f_{ \rm gas}$.}. The decline in molecular gas fraction in
the stellar mass limited sample is broadly characterised by $f_{\rm
gas}\propto(1+z)^\gamma$, with $\gamma\sim1.5$--2.5, shallower than the rate
of decline of the average specific SFR in galaxies over the same epoch, which
falls off more like $(1+z)^4$ (e.g.\ Karim et al.\ 2010). This is expected,
since molecular reservoirs can be replenished with fresh material via cooling;
thus, it is the relative evolution of the cooling rate and star formation rate
that shape the evolution of $f_{\rm gas}$.

Recent semi-analytic prescriptions for galaxy formation make predictions for
the evolution of the molecular ISM. In Fig.\ 2 we show the average ${\rm
H_2+He}$ fraction in $M_\star\geq10^{10}M_\odot$ galaxies selected in halos of
various mass from the {\sc galform} model (galaxies populated within the
Millennium Simulation $\Lambda$CDM framework). This latest model (Lagos et
al.\ 2010) implements an empirically based star-formation law (Blitz \&
Rosolowsky\ 2006) to estimate the molecular gas mass. The model predicts that
at $z<3$ $f_{\rm gas}$ is lower and the rate of decline faster for more
massive halos, and at in the mass regime pertinent to the samples presented,
$10^{12}M_\odot$, evolves in a way similar to the observations. The results
are broadly consistent with other models. For example, in smoothed particle
hydrodynamic simulations, Kere\v{s} et al.\ (2005) predict that the average
accretion rate of gas onto galaxies within halos of $M_{\rm
vir}=10^{12}M_\odot$ decreases from $\dot{M}_{\rm cold}\sim20\ M_\odot\ {\rm
yr}^{-1}$ at $z=2$ to $\dot{M}_{\rm cold}\sim3\ M_\odot\ {\rm yr}^{-1}$ at
$z=0.4$. The observed evolution is also consistent with the predicted
cosmological evolution of the accretion rate at fixed halo mass in other
models (see Dutton et al.\ 2010).

The evolution of $f_{\rm gas}$ is expected to be strongly dependent on galaxy
mass, since the evolution of the global average SFR appears to be halo mass
dependent, as does the expected evolution of the gas accretion rate. Cooling
and star formation are the key drivers of the molecular gas fraction; but it
is also sensitive to the merger rate (which can deliver molecular material,
trigger star-formation and reconfigure the baryonic content), feedback
(heating or ejection of cold gas and H$_2$ dissociation), environmental
effects (e.g.\ ram-pressure stripping) and growth in size of galaxy disks. The
latter has an important effect since it (on average) reduces the disk
hydrostatic pressure ($\propto r^2$) and therefore the efficiency of H$_2$
formation (reflected in the ${\rm H_2}$/H{\sc i} ratio, Elmegreen\ 1993; Wong
\& Blitz\ 2002), as can the metallicity and strength of the interstellar UV
radiation field.

Selection effects will affect Fig.\ 2 severely, and we should discuss how
these might affect our interpretation. Sensitivity limits do not allow us to
detect galaxies with very low gas fractions at high-$z$, and in low-$z$
surveys the very gas rich systems appear to be rare. Here we have attempted to
compare similar galaxies across a range of epochs, but the current sample
sizes are small, and inhomogeneously selected; any conclusions we draw from
Fig.\ 2 should be treated as indicative for future surveys that can more
precisely control selection and have the efficiency to survey larger numbers.

A further caveat to note is that both Daddi et al.\ (2010) and Tacconi et al.\
(2010) apply corrections to their observed CO luminosities on account of the
difference in the Rayleigh-Jeans brightness temperature of the higher {\it J}
transitions targeted. Tacconi et al. (2010)\ apply $R_{31}=0.5$ to their
CO\,($J$$=$3$\rightarrow$2) luminosities (defined $R_{\rm 31} = L'_{\rm
CO(3\rightarrow2)} / L'_{\rm CO(1\rightarrow 0)}$), and Daddi et al.\ (2010)
apply $R_{21}=0.6$ to their CO\,($J$$=$$2$$\rightarrow$1) luminosities (see
also Dannerbauer et al.\ 2009). Leroy et al.\ (2008) also apply $R_{21}=0.8$
to a sub-sample of the $z\sim0$ galaxies that were mapped in
CO\,($J$$=$$2$$\rightarrow$1). Furthermore, Daddi et al.\ (2010) assume an
$\alpha=3.6\pm0.8$, inferred from dynamical mass arguments. Although this is
within $\sim$1$\sigma$ of the Galactic value, assuming $\alpha=4.6$ increases
the gas fraction by $\sim$12\%. These systematic uncertainties in CO--H$_2$
conversions should be taken as a corollary when interpreting Fig.\ 2, however
a large change in the gas mass (or stellar mass) would be required to remove
the trend in $f_{\rm gas}$. It is unlikely that we have over-estimated the gas
mass in the $z=0.4$ LIRGs, since we have applied $\alpha=4.6$. In the high-$z$
sources, the gas masses would have to be over-estimated by a factor $\sim$5,
or the stellar masses under-estimated by a factor $\sim$4 to remove the trend.

\section{Summary}

We have presented new constraints on the molecular gas mass of $z=0.4$ LIRGs,
showing strong evolution of $f_{\rm gas}$ since $z\sim2$. The observed trend
encodes critical information on the evolution of the growth of galaxies,
including gas accretion, feedback and stellar mass assembly. Future surveys
that are able to split samples based on environment, mass, and other
parameters will disentangle the relative effects that shape the evolution of
the gas fraction. These observations will soon be supplemented and surpassed
by progress made with the Atacama Large Millimeter Array and the Expanded Very
Large Array, which will provide unprecedented sensitivity and resolution with
which to probe the evolution of gas in galaxies.

This work was based on observations made with the IRAM PdBI, supported by
INSU/CNRS (France), MPG (Germany) and IGN (Spain). We thank the anonymous
referee for a constructive report. It is a pleasure to thank Roberto Neri for
his support in this project. J.E.G. acknowledges the National Sciences and
Engineering Research Council of Canada, support from the endowment of the
Lorne Trottier Chair in Astrophysics and Cosmology (McGill) and the U.K.
Science and Technology Facilities Council (STFC). I.R.S. and A.C.E also
acknowledge STFC.


\begin{thebibliography}{}

\bibitem[Ref.  et al. (2011)]{ref11}{{Blanton}, M.~R. and {Roweis}, S., 2007, \aj, 133, 734}

\bibitem[Ref.  et al. (2011)]{ref11}{Boissier, S., Prantzos, N., Boselli, A., Gavazzi, G., 2003,
\mnras, 346, 1215}

\bibitem[Ref.  et al. (2011)]{ref11}{{Blitz}, L. and {Rosolowsky}, E., 2006, \apj, 650, 933}

\bibitem[Ref.  et al. (2011)]{ref11}{Blitz, L., Fukui, Y., Kawamura, A., Leroy, A., Mizuno, N., \& Rosolowsky, E.
2007, in Protostars and Planets V, ed. B. Reipurth, D. Jewitt, \& K. Keil
(Tucson, AZ: Univ. Arizona Press), 81}

\bibitem[Ref.  et al. (2011)]{ref11}{{Bruzual}, G. and {Charlot}, S., 2003, \mnras, 344, 1000}

\bibitem[Ref.  et al. (2011)]{ref11}{Coppin, K.\ E., K., et al.\ 2007, \apj, 665, 936}

\bibitem[Ref.  et al. (2011)]{ref11}{Czoske, O., Kneib, J.-P., Soucail, G., Bridges, T. J.,
   Mellier, Y., Cuillandre, J.-C., 2001, \aap, 372, 391}

\bibitem[Ref.  et al. (2011)]{ref11}{Daddi, E., et al., 2008, \apjl, 673, L21}

\bibitem[Ref.  et al. (2011)]{ref11}{Daddi, E., et al., 2010, \apj, 713, 686}


\bibitem[Ref.  et al. (2011)]{ref11}{Dannerbauer, H., Daddi, E., Riechers, D. A., Walter, F.,
   Carilli, C. L., Dickinson, M., Elbaz, D., Morrison, G. E.,
   2009, \apjl, 698, 178}

\bibitem[Ref.  et al. (2011)]{ref11}{Downes, D., Solomon, P.\ M., Radford,\ S.\ J.\ E., 1993,
   \apjl, 414, L13}


\bibitem[Ref.  et al. (2011)]{ref11}{Dutton, A. A., van den Bosch, F. C. Dekel, A., 2010, \mnras, 405,
1690}

\bibitem[Ref.  et al. (2011)]{ref11}{{Elmegreen, B.~G., 1993, \apj, 338, 178}}

\bibitem[Ref.  et al. (2011)]{ref11}{Fukugita, M., Hogan, C. J., Peebles, P. J. E., 1998, \apj, 503, 518}
 \bibitem[Ref.  et al. (2011)]{ref11}{Geach, J. E., et al.\ 2006, \apj, 649, 661}

\bibitem[Ref.  et al. (2011)]{ref11}{{Geach}, J.~E., {Smail}, I., {Coppin}, K., {Moran}, S.~M.,
       {Edge}, A.~C., {Ellis}, R.~S., 2009, \mnras, 395, L62 (G09)}

\bibitem[Ref.  et al. (2011)]{ref11}{{Geach}, J.~E., {Smail}, I., {Moran}, S.~M., {Treu}, T.,
               {Ellis}, R.~S., 2009, \apj, 691, 783}

\bibitem[Ref.  et al. (2011)]{ref11}{Guilloteau, S., \& Lucas, R., 2000, ASPC conference
   proceedings, 217, 299}

\bibitem[Ref.  et al. (2011)]{ref11}{{Harris}, A.~I., {Baker}, A.~J., {Zonak}, S.~G., {Sharon}, C.~E., 
{Genzel}, R., {Rauch}, K., {Watts}, G., {Creager}, R., 2010, \apj, 723, 1139}

\bibitem[Ref.  et al. (2011)]{ref11}{Ivison, R. J., Papadopoulos, P. P., Smail, Ian, Greve, T. R., Thomson, A. P., Xilouris, E. M., Chapman, S. C., 2010, astro-ph/1009.0749} 

\bibitem[Ref.  et al. (2011)]{ref11}{Karim, A., et al.\ 2010, astro-ph/1011.6370}

\bibitem[Ref.  et al. (2011)]{ref11}{{Kere{\v s}}, D., {Katz}, N., {Weinberg}, D.~H., {Dav{\'e}}, R., 2005, \mnras, 363, 2}

 \bibitem[Ref.  et al. (2011)]{ref11} {{Kennicutt}, R. C., 1998,
  \apj, 498, 541}


\bibitem[Ref.  et al. (2011)]{ref11}{Komugi, S., Yasui, C., Kobayashi, N., Hatsukade, B., Kohno, K.,
Sofue, Y., Kyu, S., 2010, PASJ in press, astro-ph/1011.3385}

\bibitem[Ref.  et al. (2011)]{ref11}{{Kormendy}, J. and {Kennicutt}, Jr., R.~C., 2004, \araa, 42, 603}

\bibitem[Ref.  et al. (2011)]{ref11}{Lagos, C., Lacey, C.\ G., Baugh, C.\ M., Bower, R.\ G., Benson, A. J., 2010, astro-ph/1011.5506}

\bibitem[Ref.  et al. (2011)]{ref11}{{Leroy}, A.~K., {Walter}, F., {Brinks}, E., {Bigiel}, F.,
       {de Blok}, W.~J.~G., {Madore}, B., {Thornley}, M.~D., 2008, \aj, 136, 2782}

\bibitem[Ref.  et al. (2011)]{ref11} {Moran}, S.M., et al., 2007, \apj, 671,
1503


\bibitem[Ref.  et al. (2011)]{ref11}{{Peng}, C.~Y., {Ho}, L.~C.,  {Impey}, C.~D.,{Rix}, {H.-W.}, 2010, \aj, 139, 2097}

\bibitem[Ref.  et al. (2011)]{ref11}{{Riechers}, D.~A., {Carilli}, C.~L., {Walter}, F., {Momjian}, E.,
2010, astro-ph/1010.4299}

\bibitem[Ref.  et al. (2011)]{ref11}{{Sakamoto}, K., {Okumura}, S.~K., {Ishizuki}, S., {Scoville}, N.~Z., 1999, \apj, 525, 691}

\bibitem[Ref.  et al. (2011)]{ref11}{{Sheth}, K., {Vogel}, S.~N., {Regan}, M.~W., {Thornley}, M.~D.,
       {Teuben}, P.~J., 2005, \apj, 632, 217}

\bibitem[Ref.  et al. (2011)]{ref11}{Solomon, P.\ M., \& Vanden\ Bout, 2005, \araa, 43, 677}

\bibitem[Ref.  et al. (2011)]{ref11}{{Solomon}, P.~M., {Downes}, D., {Radford}, S.~J.~E.,
       {Barrett}, J.~W., \apj, 1997, 478, 144}

\bibitem[Ref.  et al. (2011)]{ref11}{Tacconi, L.~J., et al.\ 2008, \apj, 680, 246}

\bibitem[Ref.  et al. (2011)]{ref11}{Tacconi, L.~J., et al.\ 2010, \nat, 463, 781}

\bibitem[Ref. et al. (2011)]{ref11}{{Treu}, T., {Ellis}, R.~S., {Kneib},
J.-P., {Dressler}, A., {Smail}, I., {Czoske}, O., {Oemler}, A., {Natarajan},
P. }, 2003, \apj, 591, 53

\bibitem[Ref.  et al. (2011)]{ref11}{Wong, T., \& Blitz, L., 2002, \apj, 569, 157}
\end{thebibliography}
\end{document}